\documentclass[manuscript=article]{achemso}

\usepackage{amsmath}
\usepackage{booktabs}
\usepackage{graphicx}
\usepackage{microtype}
\usepackage{xcolor}
\usepackage{array}
\usepackage{caption}
\usepackage{placeins}
\usepackage{xurl}

\newcommand{\DataSummaryTable}{%
\begin{table}[!htbp]
\centering
\caption{Prepared ChEMBL 36 activity dataset and fold-0 split.}
\label{tab:data-summary}
\begin{tabular}{lr}
\toprule
Quantity & Value \\
\midrule
Raw activity rows queried & 3,501,718 \\
Canonical molecules after filtering and deduplication & 1,369,515 \\
Binary activity threshold & pChEMBL $\geq$ 7.0 \\
Active molecules & 520,188 \\
Temporal holdout molecules & 274,541 \\
Fold-0 training molecules & 875,979 \\
Fold-0 internal-validation molecules & 218,995 \\
\bottomrule
\end{tabular}
\end{table}
}

\newcommand{\ModelPerformanceTable}{%
\begin{table}[!htbp]
\centering
\caption{Activity-model performance on ChEMBL 36 holdouts.}
\label{tab:model-performance}
\begin{tabular}{lrrrr}
\toprule
Split & RMSE & $R^2$ & Spearman & EF@1\% \\
\midrule
Internal validation & 0.8466 & 0.5992 & 0.7674 & 2.7331 \\
Temporal holdout & 1.1615 & 0.2473 & 0.5171 & 2.4359 \\
\bottomrule
\end{tabular}
\end{table}
}

\newcommand{\BaselineSanityTable}{%
\begin{table}[!htbp]
\centering
\caption{Selected baseline sanity checks, not a full benchmark suite. Random forest and Chemprop use sampled ChEMBL 36 splits for context only; they are not full-dataset, same-protocol baselines. The current model and previous so\_f4 comparison use the full future prediction artifacts.}
\label{tab:baselines}
\begin{tabular}{llrrrr}
\toprule
Model & Evaluation rows & RMSE & $R^2$ & Spearman & EF@1\% \\
\midrule
Current 2D MLP & Future full (274,541) & 1.1615 & 0.2473 & 0.5171 & 2.4359 \\
Previous so\_f4 & Future full (274,541) & 1.1914 & 0.2080 & 0.4806 & 2.1008 \\
Random forest Morgan & Future sample (10,000) & 1.2236 & 0.1736 & 0.4203 & 1.9859 \\
Chemprop D-MPNN & Future sample (10,000) & 1.2164 & 0.1834 & 0.4337 & 1.8100 \\
\bottomrule
\end{tabular}
\end{table}
}

\newcommand{\BaceOverlapTable}{%
\begin{table}[!htbp]
\centering
\caption{BACE external-source performance before and after fold-0 training-overlap control. Primary interpretation should use the strict scaffold-disjoint row.}
\label{tab:bace-overlap}
\begin{tabular}{lrrrrr}
\toprule
Subset & $n$ & ROC AUC & RMSE & Spearman & EF@1\% \\
\midrule
Full, provenance only & 1513 & 0.8304 & 0.9330 & 0.6958 & 2.0527 \\
Structure-disjoint & 1192 & 0.8035 & 1.0052 & 0.6641 & 2.1286 \\
Scaffold-disjoint & 962 & 0.7626 & 1.0370 & 0.6047 & 2.0253 \\
\bottomrule
\end{tabular}
\end{table}
}

\newcommand{\EgfrOverlapTable}{%
\begin{table}[!htbp]
\centering
\caption{EGFR/CHEMBL203 label-hidden operational sensitivity after fold-0 training-overlap control. This same-source replay is not independent external validation or a natural prospective library.}
\label{tab:egfr-overlap}
\begin{tabular}{lrrrrr}
\toprule
Subset & $n$ & ROC AUC & Spearman & EF@1\% & Hit@10\% \\
\midrule
Full, workflow provenance & 500 & 0.9582 & 0.7001 & 2.0000 & 1.0000 \\
Structure-disjoint & 330 & 0.9754 & 0.7479 & 2.0000 & 1.0000 \\
Scaffold-disjoint & 223 & 0.9769 & 0.7773 & 2.1038 & 1.0000 \\
\bottomrule
\end{tabular}
\end{table}
}

\newcommand{\DecisionReplayTable}{%
\begin{table}[!htbp]
\centering
\caption{Strict BACE decision-layer replay on a 100-molecule virtual batch. The risk-aware order changes the front of the list rather than uniformly dominating raw activity sorting.}
\label{tab:decision-replay}
\small
\begin{tabular}{lrrrr}
\toprule
Order & Hit@10 & Mean pIC50@10 & Hit@20 & Mean pIC50@20 \\
\midrule
Raw activity & 1.0000 & 8.2303 & 0.9000 & 7.6705 \\
Decision (triage $\rightarrow$ priority score) & 0.9000 & 7.9546 & 0.8500 & 7.6392 \\
\bottomrule
\end{tabular}
\end{table}
}

\newcommand{\OperationalABTable}{%
\begin{table}[!htbp]
\centering
\caption{Retrospective operational A/B controls on the BACE scaffold-disjoint subset. Random and scaffold-diversity rows are averaged over five seeds.}
\label{tab:operational-ab}
\begin{tabular}{llrr}
\toprule
Top N & Selection rule & Hit rate & Mean pIC50 \\
\midrule
10 & Ranking score & 1.0000 & 8.3516 \\
10 & Random & 0.4400 $\pm$ 0.1342 & 6.6667 $\pm$ 0.3905 \\
10 & Scaffold diversity & 0.4200 $\pm$ 0.1304 & 6.2401 $\pm$ 0.2620 \\
50 & Ranking score & 1.0000 & 8.2026 \\
50 & Random & 0.5240 $\pm$ 0.0654 & 6.8378 $\pm$ 0.1919 \\
50 & Scaffold diversity & 0.4400 $\pm$ 0.0678 & 6.4208 $\pm$ 0.1913 \\
100 & Ranking score & 0.8900 & 7.8911 \\
100 & Random & 0.4740 $\pm$ 0.0541 & 6.6163 $\pm$ 0.1467 \\
100 & Scaffold diversity & 0.4440 $\pm$ 0.0483 & 6.4902 $\pm$ 0.1531 \\
\bottomrule
\end{tabular}
\end{table}
}

\newcommand{\ErrorCasesTable}{%
\begin{table}[!htbp]
\centering
\caption{High-ranked inactive cases from the BACE scaffold-disjoint subset. Hashes are reported instead of full structures in the manuscript; full public benchmark rows remain in the reproducibility artifact.}
\label{tab:error-cases}
\small
\begin{tabular}{lrrrr}
\toprule
Case & Rank & True pIC50 & Predicted pIC50 & Strict-subset scaffold frequency \\
\midrule
BACE\_SD\_052 & 52 & 6.8239 & 7.7409 & 3 \\
BACE\_SD\_057 & 57 & 5.4401 & 7.7257 & 1 \\
BACE\_SD\_060 & 60 & 6.7595 & 7.7065 & 41 \\
\bottomrule
\end{tabular}
\end{table}
}

\newcommand{\StrategyDifferenceTable}{%
\begin{table}[!htbp]
\centering
\caption{Strategy-difference summary on the BACE scaffold-disjoint subset and the strict 100-molecule BACE replay. Top-10 overlap compares frozen selection rules on the same scaffold-disjoint subset; the bucket panel summarizes the label-hidden decision replay and is not a prospective risk-distribution claim.}
\label{tab:strategy-difference}
\footnotesize
\setlength{\tabcolsep}{3pt}
\begin{tabular}{>{\raggedright\arraybackslash}p{0.38\linewidth}p{0.19\linewidth}rr}
\toprule
Item & Context/share & Overlap/hit & Jaccard/pIC50 \\
\midrule
\multicolumn{4}{l}{\textit{Selection-rule contrast}} \\
\midrule
Activity rank vs. Scaffold diversity & BACE SD & 0 & 0.0000 \\
Activity rank vs. Random seed 2026 & BACE SD & 0 & 0.0000 \\
Random seed 2026 vs. Scaffold diversity & BACE SD & 0 & 0.0000 \\
\midrule
\multicolumn{4}{l}{\textit{Decision buckets in strict 100-molecule replay}} \\
\midrule
Priority & 0.1100 & 0.9091 & 8.0637 \\
Watch & 0.1800 & 0.8333 & 7.3748 \\
Low & 0.7100 & 0.4225 & 6.1861 \\
Review & 0.0000 & -- & -- \\
\bottomrule
\end{tabular}
\end{table}
}

\newcommand{\DecisionMatrixTable}{%
\begin{table}[!htbp]
\centering
\caption{Decision matrix used for review guidance. The matrix is a reporting abstraction, not an automatic experimental decision.}
\label{tab:decision-matrix}
\begin{tabular}{lll}
\toprule
Activity evidence & Domain/uncertainty evidence & Suggested review action \\
\midrule
High & Narrow interval or in-domain & First review priority \\
High & Wide interval or out-of-domain & Risk review before advancement \\
Moderate/low & Narrow interval or in-domain & Reserve or diversity control \\
Moderate/low & Wide interval or out-of-domain & Defer without external support \\
\bottomrule
\end{tabular}
\end{table}
}

\newcommand{\CostFramingTable}{%
\begin{table}[!htbp]
\centering
\caption{Assay-budget framing example for library compression. The values illustrate accounting under a fixed Top-10\% budget and are not a demonstrated project-specific savings claim.}
\label{tab:cost-framing}
\begin{tabular}{lr}
\toprule
Quantity & Value \\
\midrule
Initial library size & 10,000 \\
Compression fraction & 10\% \\
First-round tests after compression & 1,000 \\
Initial tests avoided & 9,000 \\
BACE scaffold-disjoint active recall at Top 10\% & 0.1789 \\
\bottomrule
\end{tabular}
\end{table}
}

\newcommand{\VirtualBatchTable}{%
\begin{table}[!htbp]
\centering
\caption{Virtual-batch ranking at batch size 1,000 across five library-shuffle seeds. Variability reflects batch composition, not model retraining.}
\label{tab:batch-ranking}
\small
\setlength{\tabcolsep}{4pt}
\begin{tabular}{lrrrr}
\toprule
Split & EF@1\% & EF@5\% & EF@10\% & NDCG@10 \\
\midrule
Internal validation & 2.7326 $\pm$ 0.0027 & 2.6789 $\pm$ 0.0013 & 2.6108 $\pm$ 0.0007 & 0.9510 $\pm$ 0.0000 \\
Temporal holdout & 2.4398 $\pm$ 0.0080 & 2.1513 $\pm$ 0.0035 & 1.9534 $\pm$ 0.0012 & 0.8811 $\pm$ 0.0002 \\
BACE scaffold-disjoint & 2.0253 $\pm$ 0.0000 & 2.0253 $\pm$ 0.0000 & 1.7956 $\pm$ 0.0000 & 0.9143 $\pm$ 0.0000 \\
\bottomrule
\end{tabular}
\vspace{2pt}
\begin{minipage}{0.92\linewidth}\footnotesize For BACE scaffold-disjoint ($n=962$), the batch-size 1,000 row corresponds to one near-complete virtual batch under each shuffle seed; the zero standard deviation is therefore a consequence of discrete binning rather than a formatting artifact.\end{minipage}
\end{table}
}

\newcommand{\BootstrapTable}{%
\begin{table}[!htbp]
\centering
\caption{Paired bootstrap 95\% intervals over frozen prediction rows. These intervals quantify fixed-prediction uncertainty and exclude retraining variability.}
\label{tab:bootstrap}
\begin{tabular}{lrrrr}
\toprule
Dataset & Rows & RMSE & Spearman & ROC AUC \\
\midrule
Internal validation & 218,995 & 0.8432--0.8497 & 0.7653--0.7696 & 0.8838--0.8867 \\
Temporal holdout & 274,541 & 1.1585--1.1648 & 0.5143--0.5197 & 0.7513--0.7550 \\
BACE scaffold-disjoint & 962 & 0.9861--1.0895 & 0.5598--0.6452 & 0.7315--0.7906 \\
EGFR scaffold-disjoint & 223 & 2.1179--2.3440 & 0.7306--0.8123 & 0.9596--0.9899 \\
\bottomrule
\end{tabular}
\end{table}
}

\newcommand{\ConformalTable}{%
\begin{table}[!htbp]
\centering
\caption{Split-conformal interval coverage.}
\label{tab:conformal}
\begin{tabular}{lrrr}
\toprule
Split & 90\% coverage & 80\% coverage & 90\% width \\
\midrule
Calibration & 0.9000 & 0.8000 & 2.7686 \\
Temporal holdout & 0.7799 & 0.6423 & 2.7686 \\
\bottomrule
\end{tabular}
\end{table}
}

\title{Budget-Constrained Compound Library Prioritization with Risk Awareness and Uncertainty Quantification}
\author{Shengyao Liang}
\email{pikeshuaiwe@gmail.com}
\affiliation{Independent Researcher}
\keywords{compound library compression; risk-aware ranking; conformal prediction; ADMET; cheminformatics; experimental prioritization}

\begin{document}

\begin{abstract}
\textbf{Background:} The bottleneck addressed here is not access to molecular
structures, but the limited number of compounds that can be tested, purchased,
reviewed or synthesized first. In many early projects, the immediate question
is not whether every molecule can receive a model score, but which small
fraction of a large library deserves the next unit of experimental attention.
\textbf{Objective:} I formulate this as risk-aware, budget-constrained
compound-library compression. Given a molecular library and a testing budget,
the task is to return an enriched Top-$k$ candidate set while preserving the
uncertainty, applicability-domain and risk evidence needed for human review.
\textbf{Methods:} The framework is intentionally conservative in its modeling
choices and stricter in its evidence handling. It combines a transparent 2D
activity proxy, split-conformal uncertainty intervals, applicability-domain
evidence, ADMET and structural alerts, leakage auditing, and auditable
Top-$k$ export. I use Morgan fingerprints, RDKit descriptors and a multilayer
perceptron as a fast, reviewable baseline because an upstream triage layer
should first be scalable, inspectable and difficult to over-interpret.
\textbf{Results:} On ChEMBL 36, the model achieved Spearman 0.7674 and
EF@1\% 2.7331 on internal validation, and Spearman 0.5171 with EF@1\% 2.4359
on a temporal holdout. After fold-0 training-overlap control, the
scaffold-disjoint BACE subset retained ROC AUC 0.7626 and EF@1\% 2.0253. In
a strict 100-molecule BACE decision-layer replay, the risk-aware order kept
Hit@10 at 0.9000 while exposing risk and uncertainty evidence that pure
activity sorting omits. An EGFR/CHEMBL203 label-hidden operational replay
supported workflow feasibility but is reported as same-source sensitivity
analysis rather than independent external validation.
\textbf{Conclusion:} The evidence supports risk-aware library compression as
an experimental prioritization layer before existing screening, CADD and
wet-lab workflows. The claim is intentionally bounded: retrospective
label-hidden replay is separated from truly prospective unknown-activity
library use, and externally blinded or prospective validation remains
necessary before claiming project-specific hit-rate or cost improvements.
\end{abstract}

\section{Introduction}
The motivation for this work came from a practical gap that became increasingly
clear while building a screening platform: producing a prediction for every
molecule is not the same as helping a project decide which molecules should be
tested next. Compound screening workflows commonly begin with a large
candidate space and a much smaller first-round budget. A project may have
access to hundreds, thousands, or millions of candidate structures, while only
a limited subset can enter biochemical screening, phenotypic screening,
docking, medicinal-chemistry review, purchase, or synthesis. The useful
question is therefore an upstream one: before expensive downstream work begins,
can a library be compressed into a smaller, risk-annotated candidate set whose
priority order is worth reviewing?

This paper is written from that decision point. I do not present the framework
as a replacement for high-throughput screening, molecular docking, CADD
judgment, medicinal chemistry, or wet-lab validation. I present it as a way to
make the input to those stages less indiscriminate. For screening teams in
pharmaceutical, CRO, academic and compound-supplier settings, the practical
value is a transparent pre-screening layer that reduces the first-round search
space and exposes risk evidence before compounds consume experimental budget.

Traditional quantitative structure--activity relationship (QSAR) models are
typically designed for target-specific absolute activity prediction, such as
regression of pIC50 values for a single kinase. My previous target-specific
JAK2 work \cite{Liang2025JAK2} showed that optimized machine-learning models
can capture structure--activity signal in a single-target setting. The present
pre-screening problem is different: in early
budget-constrained screening, the immediate decision need is often not precise
absolute activity against a known target, but a transparent and target-free
prioritization of a structurally diverse library. This motivates a shift from
single-target activity regression to risk-aware ranking and library compression.

Accordingly, this paper studies a target-free pre-screening layer that uses
molecular structure to rank candidates, flag uncertainty and risk, and export
Top-$k$ subsets such as the top 10\%, 5\%, 1\%, 0.5\%, or 0.1\% of a library.
The layer is meant to precede target-specific binding prediction,
high-throughput screening, molecular docking, expert CADD and experimental
validation rather than replace them. Its output is an auditable prioritization
list, not a declaration that any molecule is experimentally active. This
distinction is important because the intended collaboration is also different:
the system should help a partner decide what to inspect or test first, and the
partner's historical labels or prospective experiments should then determine
whether the enrichment is real in that project context.

The contributions of this work are:
\begin{enumerate}
  \item a budget-constrained formulation of risk-aware compound-library
        compression before experimental screening;
  \item a reproducible evaluation workspace with leakage audits for exact
        standardized-structure and Bemis--Murcko scaffold overlap;
  \item an implementation coupling a strong 2D activity baseline with
        uncertainty, applicability-domain, and risk evidence;
  \item a reporting protocol that separates internal validation, temporal
        holdout, overlap-controlled external evaluation, and label-hidden
        operational replay.
\end{enumerate}

\subsection{System Design Philosophy}
The system design is deliberately conservative. I chose not to make the paper
about a new molecular representation because that would miss the decision
problem I wanted to solve. In this setting, a useful upstream layer must be
fast enough for batch use, stable enough for repeated partner runs,
transparent enough to audit, and restrained enough not to turn a model score
into a false experimental conclusion.

Three design choices follow from that view. First, the activity model is a
strong 2D baseline rather than an opaque centerpiece. Second, every reported
result is tied to a specific evidence level: internal validation, temporal
holdout, overlap-controlled external evaluation, or label-hidden operational
replay. Third, the output is a review artifact, not just a ranked vector of
scores. The implementation tests a workflow hypothesis: when a library is
ranked and compressed under fixed budget constraints, the top portion should
be enriched relative to random or diversity-only selection while uncertainty
and risk evidence remain visible for downstream review.

\section{Related Work}
Molecular property prediction has long used fixed structural fingerprints as
efficient baselines, including extended-connectivity fingerprints
\cite{ecfp}. Graph neural networks and message-passing models, including
Chemprop \cite{chemprop}, provide a widely used alternative for supervised
molecular prediction. In practical screening, however, strong 2D baselines
remain useful because they are inexpensive, reproducible and easy to
audit at large scale.

Virtual screening is commonly evaluated with ranking metrics such as
enrichment factor, hit rate and early-recognition measures
\cite{truchon_bedroc}. These metrics align with constrained experimental
budgets because the value of a model often depends more on the composition of
the first few percent of a ranked list than on global regression accuracy.
At the same time, retrospective benchmarks can be misleading when training and
test sets share structures, scaffolds, assay sources, or time periods.
Therefore, the present work treats leakage auditing as part of the method, not
as an optional post-hoc check.

The workflow also incorporates conformal prediction \cite{vovk_conformal} to
communicate uncertainty, RDKit-based descriptors and chemical processing
\cite{rdkit}, and rule-based structural-alert concepts such as PAINS
\cite{pains} and unwanted-substructure filters \cite{brenk}. These risk signals
are used as early warnings for follow-up, not as standalone experimental
toxicity or developability conclusions.

\section{Problem Formulation}
Let $\mathcal{L}=\{x_i\}_{i=1}^{N}$ denote a molecular library and let
$B \ll N$ denote the number of molecules that can be advanced into a first
experimental or expert-review round. The goal is to produce an ordered subset
$\pi_B(\mathcal{L})$ that is enriched for useful candidates while preserving
evidence needed for downstream review. For each molecule, the workflow returns:
\begin{itemize}
  \item a batch-relative activity ranking score;
  \item uncertainty and applicability-domain evidence;
  \item ADMET and structural-risk indicators;
  \item a decision tier and exportable audit fields.
\end{itemize}

The primary scientific claim is deliberately narrow: under the evaluated data
settings, the top-ranked portion of a library is enriched for labeled active
molecules relative to the library baseline. The claim is not that top-ranked
molecules are guaranteed hits, that a project-specific mechanism has been
inferred, or that experimental confirmation can be skipped.

\subsection{Evidence Levels and Unknown-Activity Use}
The workflow separates three evidence settings that are often conflated in
screening discussions. The first is train-unseen labeled evaluation, where
labels exist but rows were not used to fit the model. Internal validation,
temporal holdout and overlap-controlled BACE belong to this setting. The
second is label-hidden operational replay, where the workflow receives only
SMILES, produces a ranked candidate list, and uses withheld labels only after
ranking is complete. The EGFR/CHEMBL203 replay and the strict BACE backend
replay follow this design. The third is prospective unknown-activity library
use, where no activity labels exist at the time of ranking. In that setting
the system can still produce Top-$k$ candidates and review evidence, but true
enrichment, hit rate and cost effects can only be measured after historical
labels are revealed or new experiments are completed.

This distinction is important for external calibration. A real external
library can be processed from anonymized SMILES without target names or project
codes, but its scientific value must be quantified through blinded historical
replay or a small prospective A/B pilot rather than by assuming that public
benchmark enrichment transfers unchanged to every project context.

\subsection{Budgeted Ranking Objective}
Let $a_i$ denote a normalized activity score, $c_i$ a confidence score, $d_i$
an applicability-domain score, and let $p_{ij}$ denote risk penalties for
structural alerts, ADMET evidence, synthesis complexity and molecular obesity.
The implementation computes a policy-dependent utility
\begin{equation}
q_i =
\operatorname{clip}_{[0,1]}\left(
w_a a_i + w_c c_i + w_d d_i - \sum_j \lambda_j p_{ij}
\right).
\end{equation}
Molecules are first separated into decision buckets representing hard blocks,
high-risk evidence, warnings and no detected concern. Here, utility $q_i$ is
the continuous internal ordering variable, whereas the reported priority score
is a review-facing 0--100 value after bucket-specific caps and soft penalties
have been applied. This distinction matters because a molecule can have a high
activity proxy and still be demoted when the risk or uncertainty evidence makes
immediate advancement questionable. Given a budget $B$, the returned candidate
set is
\begin{equation}
\pi_B(\mathcal{L}) =
\{x_i \in \mathcal{L}: \operatorname{rank}(x_i) \leq B\},
\end{equation}
where $\operatorname{rank}(x_i)$ is induced by the utility $q_i$, the decision
bucket $b_i$ and the attached review evidence $r_i$. The objective is not to
maximize a hidden scalar alone, but to produce a Top-$B$ list whose ordering
and demotions can be inspected.

This formulation separates two questions that are often conflated. The
activity proxy asks which molecules should appear near the top of a list. The
risk layer asks which evidence should be visible before those molecules consume
experimental budget. The second question may alter the order even when the
activity model itself is unchanged.

\section{Data and Leakage Control}
\subsection{ChEMBL 36 Preparation}
The primary activity dataset is derived from ChEMBL 36 \cite{chembl36}. The
preparation pipeline filters records with available pChEMBL-like values,
retains common potency types, canonicalizes structures, aggregates duplicate
canonical SMILES by median activity, and assigns scaffold plus time splits.
The binary activity label used for ranking metrics is pChEMBL $\geq 7.0$.
The temporal cutoff is 2020. Records after the cutoff form the temporal
holdout; earlier records form five scaffold folds. Fold 0 uses 875,979
molecules for training and 218,995 for internal validation. No row cap is used
for the reported current-model internal or temporal metrics.
Table~\ref{tab:data-summary} summarizes the prepared data and fold-0 split.
\DataSummaryTable

\subsection{Dataset Roles}
Each dataset is assigned a narrow evidentiary role before metrics are
interpreted. ChEMBL 36 internal validation measures in-distribution
generalization within the prepared source. The ChEMBL 36 temporal holdout
tests time-shifted train-unseen performance on records newer than the training
cutoff. BACE is an external-source benchmark and is used for primary
overlap-controlled external evidence only after exact standardized-structure
and scaffold overlap with the fold-0 training set are removed.
EGFR/CHEMBL203 is not used as independent external evidence because it is drawn
from the same ChEMBL source family and was constructed as a balanced
extreme-activity operational replay. Its role is to test whether the operational
SMILES-only workflow preserves ranking signal when labels are hidden until
after export.

\subsection{External and Workflow Evaluation Sets}
The temporal holdout contains records after the time cutoff and is used to
simulate future deployment drift. The BACE benchmark from MoleculeNet
\cite{moleculenet} is used as an external-source benchmark. A fold-0 training
audit found 321 standardized structures from the full BACE set in the training
subset, so the full BACE result is retained only for provenance. The primary
BACE evidence is reported on structure-disjoint and scaffold-disjoint subsets.

An EGFR/CHEMBL203 label-hidden replay was built to exercise the operational
workflow. The uploaded file contains SMILES only; labels are withheld during
ranking and revealed afterward for evaluation. The formal fold-0 audit
identified 170 standardized-structure matches among 500 molecules, leaving a
structure-disjoint subset of 330 molecules and a Bemis--Murcko
scaffold-disjoint subset of 223 molecules. Because these records remain from
the same ChEMBL 36 source family, the EGFR replay is reported as a
target-specific same-source operational sensitivity analysis, not as independent
external validation. The EGFR sample is also a balanced extreme-activity set
with 250 active and 250 inactive molecules, excluding the 100--1000 nM middle
activity range; its prevalence is not representative of a natural prospective
screening library.

\subsection{Leakage-Control Hierarchy}
Every external or replay dataset is audited against the exact fold-0 training
subset, not against a generic project-wide file. Structures are standardized
with the same chemistry stack used by the evaluation workflow. Two exclusion
levels are reported:
\begin{enumerate}
  \item \textbf{standardized-structure disjointness}, which removes exact
        standardized molecular matches; and
  \item \textbf{Bemis--Murcko scaffold disjointness}, which additionally
        removes molecules whose core scaffold occurs in training.
\end{enumerate}
The full datasets are retained only as provenance records. Primary external
claims use the strictest available subset. Source provenance, time ordering
and sample construction are reported separately because zero exact overlap
does not make a same-source, target-selected replay equivalent to an
independent prospective library.

\section{Method}
\subsection{Activity Proxy}
The current activity model is a deliberately transparent 2D
baseline. I chose this design because upstream triage needs fast batch
inference, stable behavior and auditability more than architectural novelty.
Each molecule is represented by a 2058-dimensional vector: a 2048-bit Morgan
fingerprint (radius 2) concatenated with 10 RDKit descriptors. The descriptor
block includes molecular weight, logP, polar surface area, hydrogen-bond
counts, rotatable bonds, ring count, fraction sp$^3$, heavy-atom count and
Labute ASA. Descriptor features are standardized using fold-specific training
statistics, concatenated with the fingerprint vector, and passed into a
multilayer perceptron with hidden widths 1024--512--256, ReLU activations,
batch normalization and dropout. Training minimizes a weighted mean-squared
activity loss together with a pairwise ranking term, with additional emphasis
on hard-negative activity ranges near the active threshold. The model is
selected by a validation score that emphasizes Spearman correlation and early
enrichment rather than only global regression fit.

\subsection{Uncertainty, Domain Evidence and Risk Signals}
Split-conformal intervals are calibrated on the internal validation fold using
absolute residual quantiles. The output interval is interpreted as a
fixed-model uncertainty signal rather than a guarantee under arbitrary
distribution shift. Applicability-domain evidence is derived from structural
similarity and local-neighborhood behavior in the prepared activity data, so
the system can distinguish a high-scoring in-domain molecule from a
high-scoring borderline case.

ADMET and structural alerts are incorporated as review evidence. The decision
layer separates hard-blocking evidence from soft warning evidence so that a
single risk flag does not collapse all ranked candidates into the same
decision bucket. In practice, this means the front of the list can still be
ranked by activity while remaining visibly annotated for hERG, AMES, CYP,
solubility, stability and synthesis-related concerns.

\subsection{Export and Audit Layer}
The implementation produces Top-$k$ CSV/XLSX candidate sets and molecule-level
PDF reports. The PDF reports contain the same decision fields, uncertainty
intervals and risk alerts as the tabular export, formatted for single-molecule
inspection after a smaller candidate set has been selected. Exported rows keep
review-facing decision fields near the front and audit fields later in the
table. This design reflects the intended use: external users first review the
compressed candidate list and then inspect selected molecules in greater
detail. Batch export is supported at practical compression levels such as Top
20\%, 10\%, 5\%, 1\%, 0.5\% and 0.1\%, so the same workflow can serve both
large-library triage and small follow-up reviews.

\subsection{System Implementation and Operational Workflow}
The system accepts SMILES as the minimal input and does not require target
names, project identifiers or biological context for the ranking pass. The
workflow performs input quality control, standardization, duplicate handling,
batch inference, multi-objective ranking, Top-$k$ export and molecule-level
report generation. The replay implementation preserves batch-level provenance
so that exported candidate sets can be traced back to the submitted input,
model artifact and ranking policy.

Operational outputs are intentionally split into compact decision artifacts
and deeper audit artifacts. CSV/XLSX exports are used for library-level review
and downstream tooling; molecule-level PDF reports are used only after a
smaller set of candidates has been selected. This avoids turning a large batch
run into a collection of unstructured single-molecule reports.

\begin{figure}[t]
  \centering
  \includegraphics[width=0.95\linewidth]{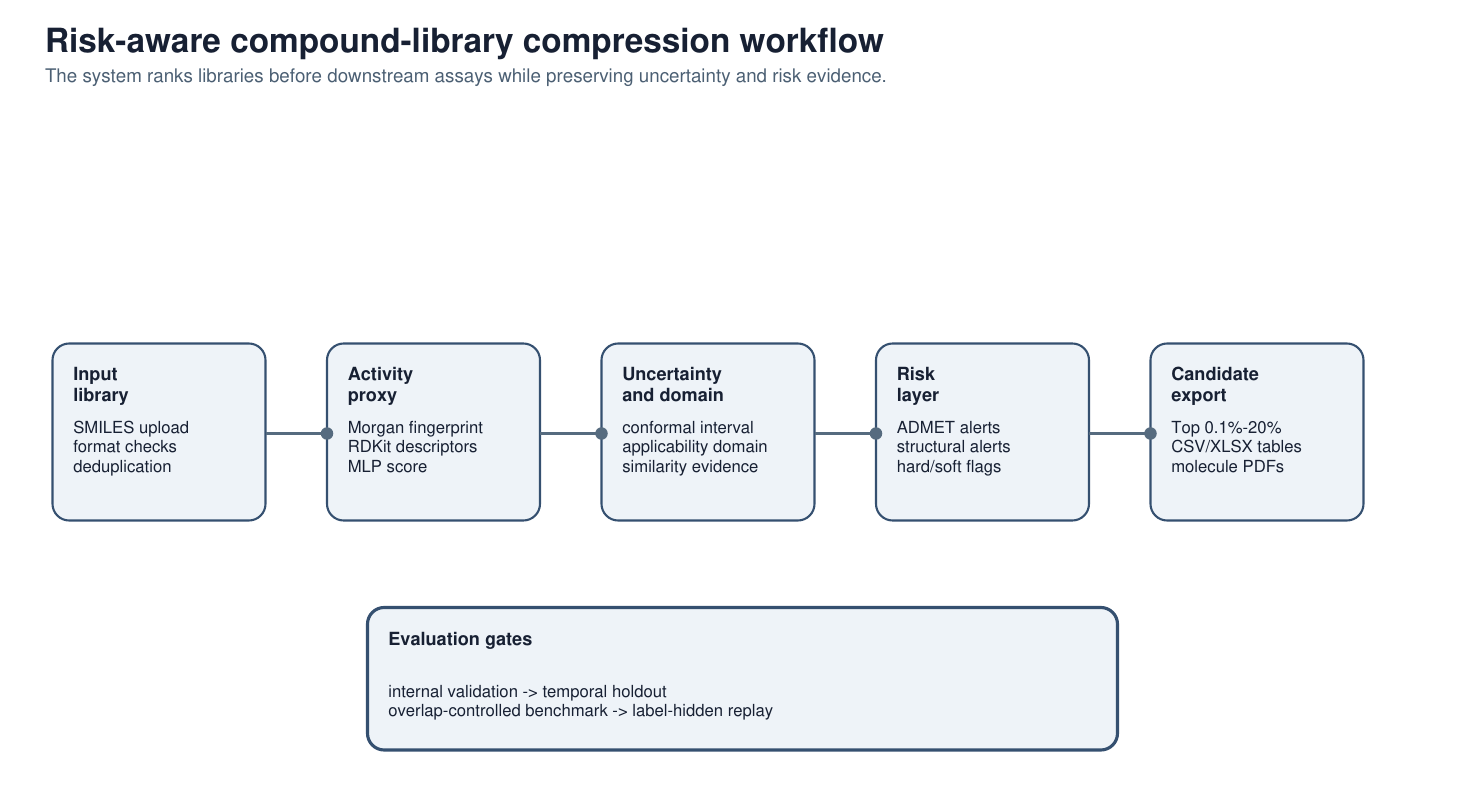}
  \caption{Overview of the risk-aware compound-library compression workflow.}
  \label{fig:workflow}
\end{figure}

\FloatBarrier
\clearpage
\section{Experimental Protocol}
Regression metrics include RMSE, $R^2$ and Spearman correlation. Binary ranking
metrics use the active threshold pChEMBL $\geq 7.0$ and include hit rate,
recall, enrichment factor and NDCG at budgeted cutoffs. For a ranked list with
top-$k$ active rate $h_k$ and full-library active prevalence $p$, enrichment
factor is $EF@k=h_k/p$.

I report three complementary evaluation layers. First, ChEMBL 36 internal
validation and temporal holdout measure how the activity proxy behaves on
frozen split artifacts. Second, overlap-controlled BACE and EGFR subsets test
how the same framework behaves after removing exact standardized-structure or
scaffold matches from training. Third, virtual-batch compression evaluates the
system at batch sizes that resemble external batch submissions rather than single-molecule
benchmarks.

Virtual-batch experiments shuffle each labeled evaluation set into contiguous
batches and evaluate Top-1\%, Top-5\% and Top-10\% ranking quality within each
batch. Five shuffle seeds are used for the main virtual-batch table. Paired
nonparametric bootstrap intervals are computed over frozen prediction rows.
These intervals quantify uncertainty in the fixed prediction artifacts and do
not include model retraining variability.

The strict BACE decision replay uses the reference implementation in
label-hidden mode: the system ranks a batch from SMILES only, exports the
candidate list, and restores labels only after the ranking path is complete.
This makes the replay useful for testing the operational path without
presenting the library as a prospective unknown-activity set.

\section{Results}
\begin{figure}[t]
  \centering
  \includegraphics[width=0.95\linewidth]{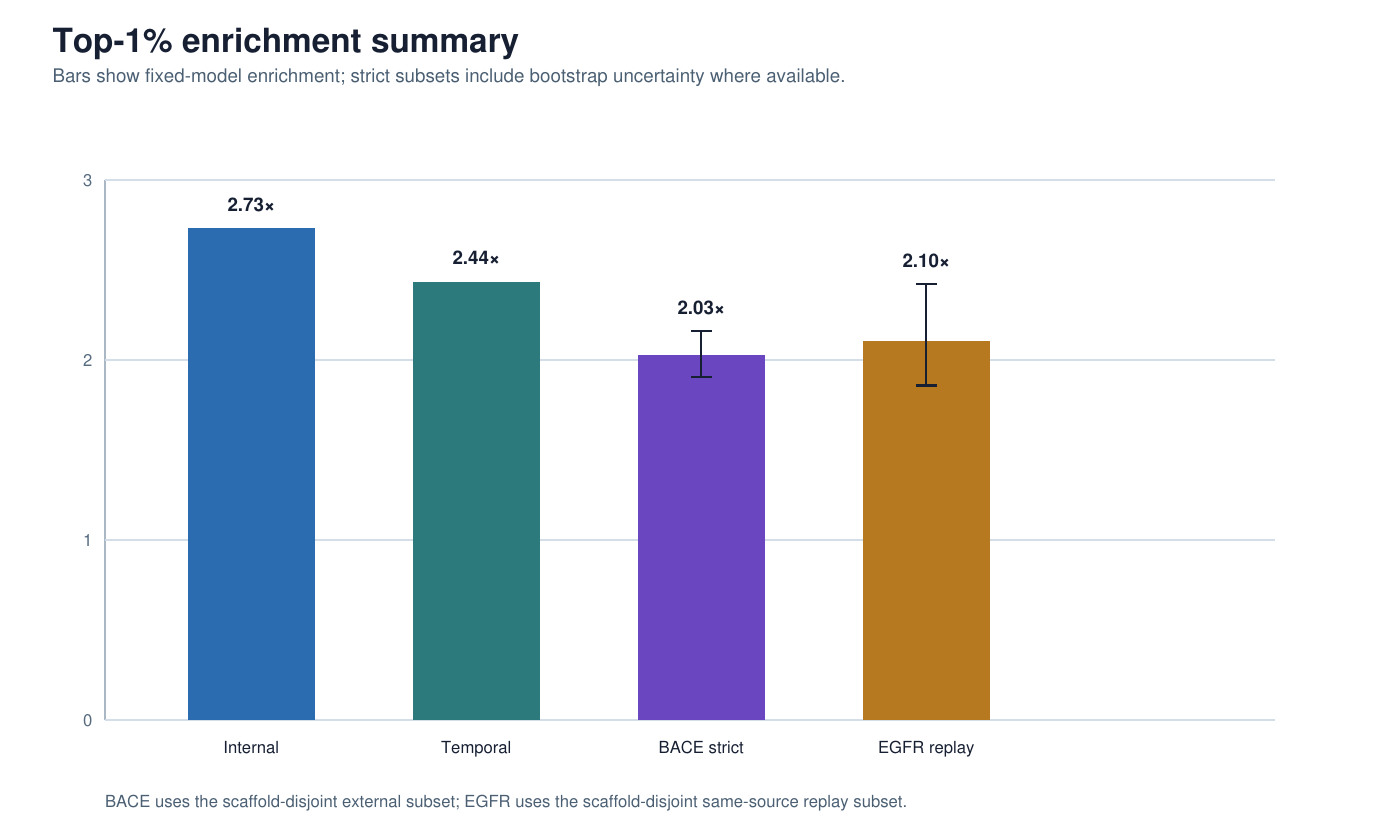}
  \caption{Top-1\% enrichment on frozen evaluation artifacts: 2.73$\times$
  for ChEMBL 36 internal validation, 2.44$\times$ for ChEMBL 36 temporal
  holdout, 2.03$\times$ for BACE scaffold-disjoint, and 2.10$\times$ for EGFR
  scaffold-disjoint replay. Error bars are shown where bootstrap intervals are
  available for the strict overlap-controlled subsets. BACE uses the
  scaffold-disjoint external subset. EGFR uses the scaffold-disjoint
  same-source replay subset and is not an independent external benchmark.}
  \label{fig:top1}
\end{figure}

Figure~\ref{fig:top1} summarizes Top-1\% enrichment across the main evidence
levels. Tables~\ref{tab:model-performance}--\ref{tab:conformal} are generated
directly from machine-readable artifacts in the reproducibility workspace.

On ChEMBL 36, internal validation shows RMSE 0.8466, Spearman 0.7674 and
EF@1\% 2.7331. Temporal holdout performance decreases, as expected under time
shift, but remains useful for ranking with Spearman 0.5171 and EF@1\% 2.4359.
The previous so\_f4 model, random forest and Chemprop checks provide context,
but only the current model and previous so\_f4 future comparison use the full
future artifacts. The sampled random forest and Chemprop rows should therefore
be interpreted as sampled model sanity checks rather than full Future
all-row, same-protocol benchmarks. I include them to make the baseline context
visible, not to claim a definitive architecture ranking. A complete
same-protocol Chemprop or graph-model benchmark would require full-dataset
training and inference under the same split, preprocessing and budgeted
ranking protocol, and is left outside the present reproducibility package.
\ModelPerformanceTable
\BaselineSanityTable

After removing fold-0 training overlap from BACE, the scaffold-disjoint subset
retains ROC AUC 0.7626, Spearman 0.6047 and EF@1\% 2.0253. This is lower than
the full BACE result, which confirms that overlap control materially changes
the strength of the external claim. The EGFR replay remains strong after
overlap removal, but it is same-source and balanced by construction; it
supports operational sensitivity rather than independent external generalization.
\BaceOverlapTable
\EgfrOverlapTable

The new strict BACE decision replay on a 100-molecule virtual batch shows the
same design trade-off at the system layer. Raw activity sorting gives
Hit@10 of 1.0 and Hit@20 of 0.9, while the risk-aware decision order gives
0.9 and 0.85 on the same batch. That is not a uniform improvement on every
metric; it is evidence that the system changes which compounds are promoted
into the front of the list, making risk and uncertainty visible before the
next experimental round.
\DecisionReplayTable
\FloatBarrier

Table~\ref{tab:strategy-difference} separates two operational questions that
are easy to conflate. The first block shows how much the Top-10 candidate set
changes when selection is driven by activity ranking, scaffold diversity or a
random control on the same BACE scaffold-disjoint artifact. The second block
summarizes the decision buckets observed in the strict 100-molecule BACE
backend replay. These values are a strategy-difference display, not a claim
that any one strategy is uniformly superior under every criterion.
\StrategyDifferenceTable
\FloatBarrier

\subsection{Operational Validation on Simulated Batches}
Operational validation asks whether a budgeted selection rule behaves
differently from simple controls under the same frozen labels. On the BACE
scaffold-disjoint subset, ranking-score selection is compared with random
selection and scaffold-diversity selection at Top 10, Top 50 and Top 100.
The controls are averaged over five seeds. These experiments do not replace
prospective validation, but they provide a practical baseline for evaluating
whether the ranking workflow is doing more than selecting arbitrary or merely
diverse structures.
\OperationalABTable

The same framing can be used for assay-budget accounting. For example, a
10,000-compound library compressed to Top 10\% would produce 1,000 first-round
tests and avoid 9,000 initial tests. Table~\ref{tab:cost-framing} reports this
as an accounting framework and anchors it to the BACE scaffold-disjoint recall
observed in the frozen retrospective artifact. It should not be interpreted as
a project-specific savings claim without external calibration.
\CostFramingTable

The decision matrix in Table~\ref{tab:decision-matrix} summarizes how activity,
uncertainty and domain evidence are combined during review. The matrix is a
reporting abstraction rather than an automatic experimental decision rule.
\DecisionMatrixTable

\begin{figure}[t]
  \centering
  \includegraphics[width=0.95\linewidth]{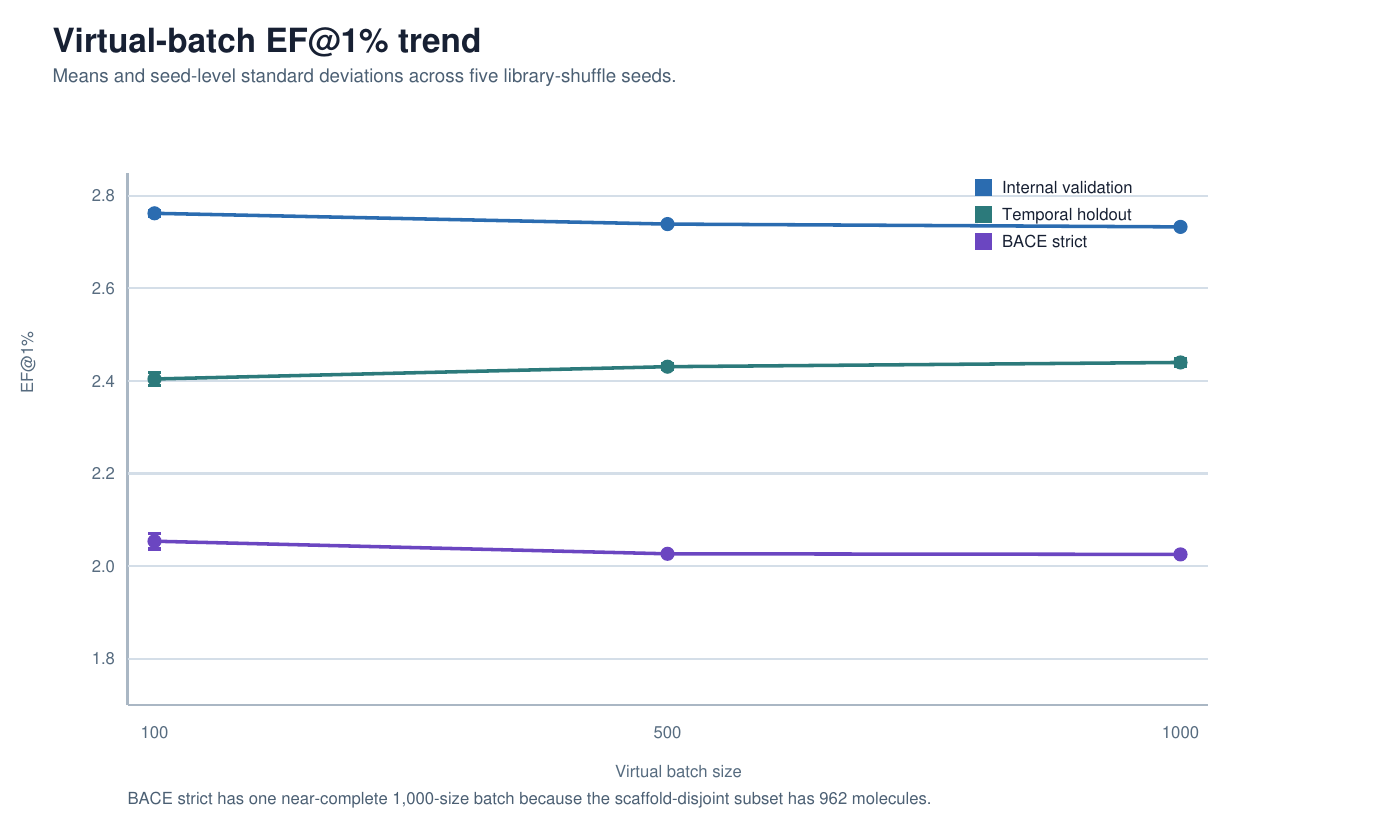}
  \caption{EF@1\% trends across virtual batch sizes. Points show means across
  five library-shuffle seeds; error bars show seed-level standard deviations.
  The BACE scaffold-disjoint row at batch size 1,000 has essentially zero
  variation because the subset contains 962 molecules, so each seed evaluates
  the same single near-complete batch.}
  \label{fig:virtual-batch-trend}
\end{figure}

Across five virtual-library shuffles, the internal, temporal and strict BACE
subsets retain useful early enrichment at a batch size of 1,000. Paired
bootstrap intervals quantify uncertainty in the frozen prediction artifacts,
while the conformal results show that nominal coverage degrades under temporal
shift. Figure~\ref{fig:virtual-batch-trend} visualizes EF@1\% across virtual
batch sizes, supporting the view that early enrichment is not an artifact of a
single batch-size choice. These analyses do not measure model-retraining
variability.
\VirtualBatchTable

\BootstrapTable
\ConformalTable
\clearpage

\section{Discussion}
The results support three practical conclusions. First, a transparent 2D model
can be strong enough to serve as the activity backbone of an early compression
layer when the task is ranking under a fixed budget. Second, Top-$k$ ranking
metrics are closer to the screening decision than global regression metrics
alone, because the experimental question is often which 1--10\% of a library
should be seen first. Third, leakage auditing is not a cosmetic addition. It
changes how external and replay results should be read, and it prevents a
useful system from being defended with evidence that is stronger than the data
allow.

The experience behind this work is that model performance and decision value
are related but not identical. A model can assign plausible scores to many
molecules and still leave a collaborator with the same operational problem:
too many compounds, too little first-round budget, and too little context about
which risks should be reviewed before testing. The most important contribution
is therefore not a claim that this particular model is final. It is the
decision framing: a large library can be compressed into a smaller, reviewable
set while preserving the evidence used to keep, demote or revisit individual
molecules. This makes the workflow suitable as an upstream layer for existing
screening, CADD, procurement and experimental stacks rather than a replacement
for them.

Risk-aware ranking also introduces a deliberate and visible trade-off. A pure
activity ranker may place every high predicted activity molecule at the top,
while the risk-aware layer can demote candidates with uncertainty,
applicability-domain or risk concerns. This is not uniformly superior to
activity-only sorting; it is useful when the decision maker values early
visibility into risk and audit evidence. In that sense, the risk layer is not
an attempt to hide model uncertainty. It is the mechanism by which uncertainty
is made operational.

For practical use, the framework should be treated as a pre-screening and
review queue rather than as an automatic selection rule. A user with a
100,000-compound library and a first-round experimental budget of 1,000 tests
would start by exporting the Top 1\% candidate set, then inspect ADMET,
structural-alert, uncertainty and applicability-domain fields before choosing
which molecules enter orthogonal testing. The Top 5--10\% lists can be kept as
backup pools for diversity review, procurement constraints, chemistry
feasibility or a second experimental round. When the library is unlabeled, I
would not ask a partner to accept the predicted score as the final evidence.
The appropriate success criterion is the subsequent enrichment of tested
compounds relative to random, diversity-matched or local-selection baselines.
This use pattern keeps the system aligned with its evidence base: it compresses
a large search space, surfaces risk earlier and preserves an audit trail, while
leaving final experimental decisions to downstream assay results and domain
review.

Figure~\ref{fig:virtual-batch-trend} adds a second operational view: early
enrichment remains stable as virtual batch size changes, but the uncertainty
around small external subsets is visibly different from the uncertainty around
large ChEMBL holdouts. This is why I separate frozen-row bootstrap,
shuffle-seed variability and future model-retraining variability instead of
collapsing them into one headline number.

\subsection{Error Analysis and Scope of Validity}
High-ranked inactive molecules are expected in any retrospective screening
artifact, especially under scaffold-disjoint evaluation. Table~\ref{tab:error-cases}
reports three such BACE cases using hashed identifiers. These cases are
interpreted as scope-boundary examples: they show where ranking evidence should
be accompanied by domain, uncertainty and risk review rather than treated as a
standalone activity conclusion.
\ErrorCasesTable

The three cases are informative rather than merely inconvenient. Two have true
pIC50 values just below the active threshold, so the binary inactive label
partly reflects thresholding near a continuous activity boundary. The third is
a stronger miss with a singleton scaffold in the strict subset, which is the
type of case where a target-free 2D prior has limited mechanistic context. The
case with a scaffold frequency of 41 also shows the opposite failure mode:
even when a scaffold is represented inside the strict BACE subset, local
analogs do not guarantee that every member has the same activity. These
examples motivate the risk-aware review layer. The ranking score proposes a
shortlist; it should not be treated as a mechanistic explanation or a final
experimental decision.

The intended interpretation is therefore not that every high-ranked molecule is
active, but that the workflow can enrich early positions while still exposing
the uncertainty and audit information needed to decide whether a molecule
should be advanced, reviewed or deferred.

\section{Limitations}
This paper reports a decision framework built on a strong baseline model
rather than a novel molecular representation. The random forest and Chemprop
rows are sampled sanity checks, not full-dataset, same-protocol baselines; they
should not be used to claim architectural superiority over graph neural
networks. Only one full training fold is currently reported, and the bootstrap
intervals do not include retraining variability. Multi-fold stability
assessment is prepared in the reproducibility workflow, but I do not report a
partial multi-fold estimate because incomplete folds would give a false sense
of precision. Public retrospective data cannot establish project-specific
hit-rate improvement or cost reduction. The EGFR replay is same-source and
balanced, and must not be described as a prospective unknown-activity library.
ADMET endpoints vary in temporal generalization; weak endpoints such as CYP
induction should be treated only as low-confidence review prompts. Prospective
or externally blinded experiments are still required for stronger deployment
claims. These limitations are not side notes to the work; they define the
level at which the evidence should be used.

\section{Data and Code Availability}
The primary activity records are derived from ChEMBL 36, and BACE is
distributed through MoleculeNet. The manuscript tables and figures are
generated from versioned machine-readable artifacts in the accompanying
reproducibility workspace. The public reproducibility repository is available
at \url{https://github.com/ShengyaoLiang/risk-aware-compound-library-compression}.
Archived releases are indexed under the Zenodo latest-record DOI
\url{https://doi.org/10.5281/zenodo.20833014}.
I release the non-sensitive source package so that the evidence chain can be
inspected without exposing operational data. It includes evaluation scripts,
dataset manifests, overlap-controlled public subset files, generated tables
and result artifacts. Runtime account data, external batch submissions,
withheld labels and credentials are excluded from the submission package.
Complete ChEMBL-derived training assets and full train/validation ID lists are
not part of the public package because of size, provenance and internal-asset
boundaries. The released artifacts are intended to reproduce the manuscript
tables from frozen non-sensitive summaries and public subset files, not to
serve as a complete service distribution.

\section{Conclusion}
Risk-aware compound-library compression offers an auditable interface between
large molecular collections and limited experimental budgets. The present
evidence supports ranking enrichment on internal validation, temporal holdout,
an overlap-controlled external-source BACE subset and a same-source
label-hidden EGFR operational replay. I therefore read the work as a case for
a pre-screening decision layer: compress first, expose risk early, and keep the
evidence trail intact before downstream experiments or expert review consume
limited resources.

The next validation step is not another uncontrolled benchmark but an
externally blinded historical replay or a small prospective A/B experiment.
In such a study, Top-$k$ candidates should be compared against random,
diversity-matched and existing project-selection baselines. If enrichment is
observed in a partner's own historical or prospective data, project-specific
hit-rate, time and assay-budget effects can then be measured directly. Until
that step is completed, the responsible claim is ranking enrichment and
decision support, not guaranteed experimental success. That boundary is not a
weakness of the platform; it is the condition under which the platform can
earn trust.

\appendix
\section{Supplementary Material: System Interface Specification}
The reference implementation accepts plain SMILES as the minimal molecular
input. Optional metadata can include a batch name, row identifier and
non-sensitive user-side labels for local traceability, but target names,
project codes and biological context are not required for the target-free
ranking pass. Standard output artifacts include:
\begin{itemize}
  \item Top-$k$ CSV/XLSX files for budgeted library review;
  \item molecule-level PDF reports for selected candidates;
  \item audit fields for ranking score, uncertainty, applicability-domain
        evidence, structural alerts and risk categories; and
  \item provenance records linking exports to batch identifiers and frozen
        software/model artifacts.
\end{itemize}

Typical integration points include workflow systems such as KNIME, Pipeline
Pilot, Spotfire-style dashboards, ELN/LIMS imports and internal data lakes. The
export contract is intentionally tabular so that downstream groups can consume
candidate lists without adopting a specific user interface. Account-level
isolation was used in the operational replay environment; an external
deployment should additionally define transport security, retention policy,
access control, logging and incident-response procedures.

\section{External Validation Protocols}
Two validation protocols are recommended for external calibration. In a
blinded retrospective replay, an external group provides SMILES while holding
back labels. The system ranks and exports Top-$k$ candidates, after which the
external group reveals labels for EF, hit-rate, recall and NDCG estimation.
This protocol is low-cost and directly tests whether the ranking layer adds
value relative to random or diversity-based controls in that data context.

In a prospective A/B pilot, candidate sets are selected by the ranking
workflow, by random or diversity-matched controls, and by an existing local
selection process if one is available. Experimental readouts are then compared
after testing. This is the appropriate protocol for project-specific hit-rate
or cost claims; such claims are not made from the retrospective public
benchmarks in this paper. This is also the collaboration path I consider most
scientifically honest: let the platform rank first, keep the labels hidden,
and allow the partner's data to decide whether the compression layer adds
value.

\section*{Author Contributions}
Shengyao Liang conceived and designed the study, prepared and evaluated the
datasets and models, implemented the reproducibility workflow, analyzed the
results, and wrote the manuscript.

\section*{Funding}
This work received no external funding.

\section*{Competing Interests}
I developed the described framework and may seek external collaborations or
pilot studies based on this work. No external funding was received, and no
third-party commercial entity influenced the study design, analysis or
manuscript.

\bibliography{references}
\end{document}